\documentclass[runningheads]{llncs}
\usepackage[T1]{fontenc}
\usepackage{amsmath}
\usepackage{amssymb}
\usepackage{graphicx,verbatim}
\usepackage{multirow}
\usepackage{xcolor,colortbl}
\usepackage{color}		
\usepackage{cite}

\definecolor{lightred}{rgb}{1,0.9,0.9}
\definecolor{lightblue}{rgb}{0.9,0.9,1.0}
\definecolor{red}{rgb}{1,0,0}
\definecolor{lightred_darker}{rgb}{1,0.7,0.7}
\definecolor{lightgreen}{rgb}{0.7,1.0,0.7}

\begin{document}
%
\title{MEGAN: Mixture of Experts for Robust Uncertainty Estimation in Endoscopy Videos}
%

\author{Damola Agbelese\inst{2}{$^{\ast}$}, Krishna Chaitanya\inst{1}{$^{\ast}$}, Pushpak Pati\inst{1}, 
Chaitanya Parmar\inst{1}, Pooya Mobadersany\inst{1}, Shreyas Fadnavis \inst{1}, Lindsey Surace\inst{1},  Shadi Yarandi\inst{1}, Louis R. Ghanem\inst{1},  Molly Lucas\inst{1}, Tommaso Mansi\inst{1}, Oana Gabriela Cula\inst{1}, Pablo F. Damasceno\inst{1}{$^{\dag}$}, Kristopher Standish\inst{1}{$^{\dag}$}}

\authorrunning{K. Chaitanya et al.}
\institute{
\textsuperscript{1} Janssen R\&D, LLC, a Johnson \& Johnson Company \\
\textsuperscript{2} ETH Zurich, Switzerland \\
\email{kchaita6@its.jnj.com}
}

\maketitle              
\begin{abstract}
\footnote {$^{\ast}$ Equal contribution as first authors. $^{\dag}$ Equal contribution as last authors.}

Reliable uncertainty quantification (UQ) is essential in medical AI. Evidential Deep Learning (EDL) offers a computationally efficient way to quantify model uncertainty alongside predictions, unlike traditional methods such as Monte Carlo (MC) Dropout and Deep Ensembles (DE). However, all these methods often rely on a single expert’s annotations as ground truth for model training, overlooking the inter-rater variability in healthcare.
To address this issue, we propose MEGAN, a Multi-Expert Gating Network that aggregates uncertainty estimates and predictions from multiple AI experts via EDL models trained with diverse ground truths and modeling strategies. MEGAN’s gating network optimally combines predictions and uncertainties from each EDL model, enhancing overall prediction confidence and calibration.
We extensively benchmark MEGAN on endoscopy videos for Ulcerative colitis (UC) disease severity estimation, assessed by visual labeling of Mayo Endoscopic Subscore (MES), where inter-rater variability is prevalent. In large-scale prospective UC clinical trial, MEGAN achieved a 3.5\% improvement in F1-score and a 30.5\% reduction in Expected Calibration Error (ECE) compared to existing methods. Furthermore, MEGAN facilitated uncertainty-guided sample stratification, reducing the annotation burden and potentially increasing efficiency and consistency in UC trials.


\end{abstract}

\section{Introduction}

Uncertainty quantification (UQ) is crucial in medical image analysis, particularly because deep learning models often produce overconfident predictions despite inherent ambiguities in clinical assessments. In practice, the absence of a universally accepted ground truth often leads to significant inter- and intra-reader variability, adversely impacting diagnoses, disease severity evaluations, and prognostic predictions. Conventional deep learning approaches typically overlook this variability, causing  inflated confidence even when clinical agreement is low. Addressing UQ is essential for automating disease severity assessments, where it is crucial to minimize subjectivity while preserving trust in AI-driven decisions.

While traditional UQ methods like MC Dropout~\cite{gal2016dropout} and Deep Ensembles~\cite{lakshminarayanan2017simple} (DE) are widely used, their post-hoc nature limits real-time deployment. MC Dropout needs multiple inference rounds, and DE methods execute multiple models in inference, both incurring high computation cost. 
In contrast, Evidential Deep Learning (EDL)~\cite{sensoy2018evidential} presents an efficient solution by estimating prediction confidence and uncertainty in a single forward pass. EDL has been applied across various domains, including MRI\cite{zou2022tbrats}, PET-CT\cite{huang2021evidential}, and X-ray\cite{ghesu2019quantifying,ghesu2021quantifying}. However, current EDL approaches typically rely on single-expert annotations during training, restricting their capacity to manage inter-rater variability, leading to model overconfidence even when experts disagree. These limitations are even more pronounced in subjective domains like endoscopy, where multiple experts manually evaluate samples, resulting in high inter-rater variability.

To address these challenges, we introduce MEGAN (Multi-Expert GAting Network), a novel framework that aggregates uncertainty estimates from multiple EDL-based models, trained on diverse expert annotations and modeling strategies. Here, each model acts as an independent AI expert with its respective strengths and weaknesses. MEGAN uses a gating network to dynamically combine their individual predictions and uncertainties, reducing inter-rater variability and enhancing overall uncertainty calibration and prediction accuracy.

We thoroughly evaluated MEGAN on endoscopy videos for Ulcerative Colitis (UC) severity estimation, specifically focusing on the Mayo Endoscopic Subscore (MES) assessment~\cite{schroeder1987coated}, which exhibits high inter-rater variability in clinical trials. UC is a chronic inflammatory bowel disease affecting $\sim$5 million individuals globally. UC severity is assessed using MES, where gastroenterologists assign severity grades on an ordinal scale (0–3) based on mucosal appearance from endoscopic videos. Accurate MES estimation is critical in UC clinical trials for patient enrollment and treatment efficacy quantification. Recently, deep learning has been widely adopted to automate MES estimation, employing fully supervised~\cite{stidham2019performance, vasilakakis2020weakly, stidham2024using}, weakly-supervised~\cite{schwab2022automatic, polat2022class, tian2022contrastive}, and self-supervised~\cite{wang2023foundation, hirsch2023self, chaitanya2024arges, dermyer2025endodino} methods. However, MES is inherently subjective, time-consuming, and prone to high inter-rater variability~\cite{rubin2023development}, requiring multiple experts to annotate the same video in trials. The existing deep learning solutions, though successful on estimating single expert MES annotations, often fail to account for inter-rater variability and prioritize accuracy over UQ. This hampers the assessment of model confidence, an essential element for practical clinical deployment.

Our key contributions in this paper are as follows: \\
\noindent 1. \textbf{Multi-expert uncertainty estimation and fusion}: We introduce MEGAN, a novel framework that aggregates estimates from multiple EDL-based models, enhancing overall uncertainty calibration and addressing inter-rater variability. \\
\noindent 2. \textbf{Large-scale UC Clinical Evaluation}: MEGAN was evaluated on multiple UC trials, achieving a 5.2\% F1-score improvement and an 29.4\% ECE reduction on Test set. On Unseen set, F1 improved by 3.5\% and ECE decreased by 30.5\%, demonstrating MEGAN's superior performance, calibration, and generalization. \\ 
\noindent 3. \textbf{Uncertainty-guided UC Analysis}: On an unseen set, MEGAN can reduce the overall number of videos to be reviewed by experts by 10\% compared to baselines. 
Further, in a subset reviewed by 3 experts, MEGAN distinguished confident cases with 14\% higher F1 than consensus rating; and successfully identified difficult cases for selective expert review.



\section{Methods}
\begin{figure}[!t]
    \centering
    \includegraphics[width=0.9\linewidth]{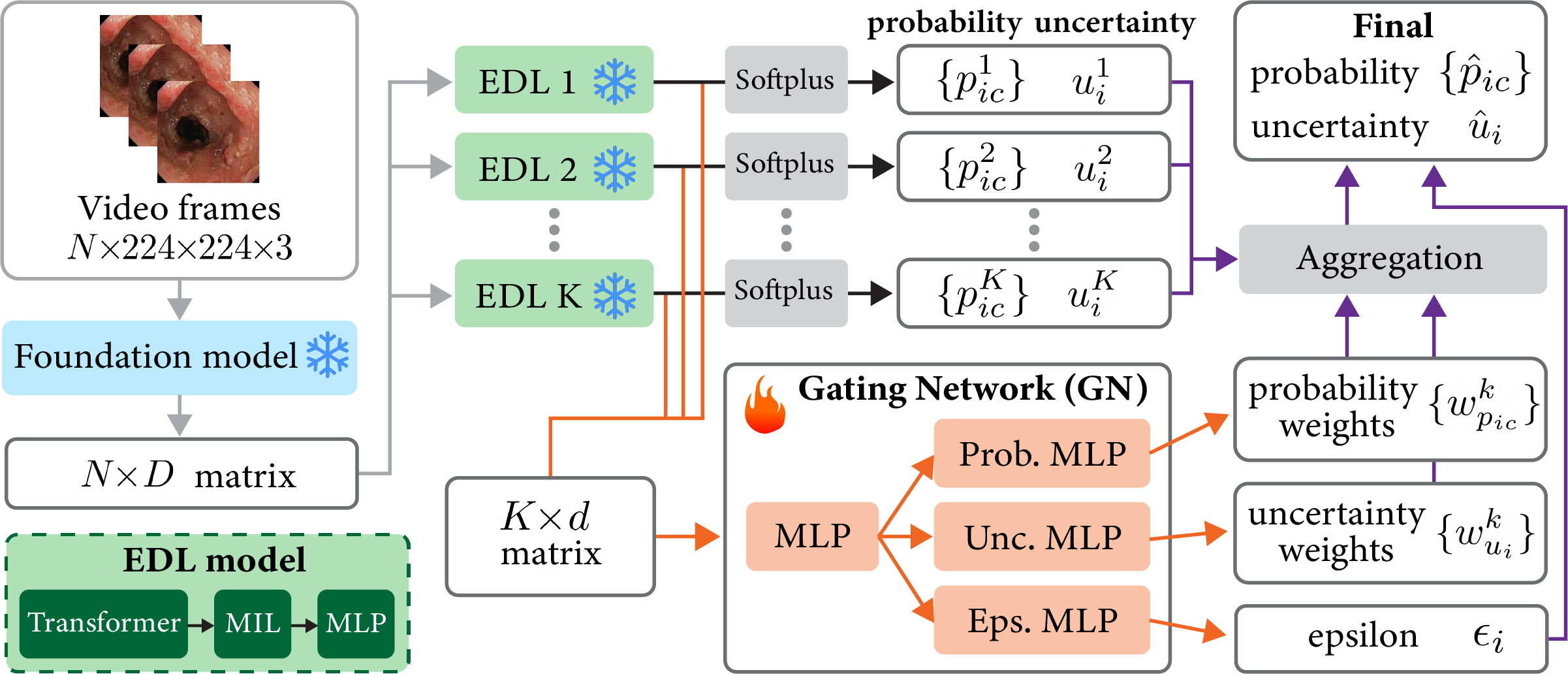}
    \caption{Overview of MEGAN architecture variants. MEGAN-Gated uses a trainable gating network (GN) to optimally combine predictions from multiple pre-trained EDL models to output a final prediction and uncertainty score.} 
    \label{fig:MEGAN_framework}
\end{figure}

We begin this section by outlining MES scoring process (Sec 2.1), followed by describing EDL expert model (Sec 2.2). Finally, we introduce MEGAN (Sec 2.3), which aggregates multiple EDL model outputs via a gating network to improve MES scoring accuracy and simultaneous uncertainty estimation (Fig. 1).

\subsection{MES Scoring Process}
In clinical trials, MES labels (integers, 0–3) are assigned through a multi-step review to ensure consistency and minimize variability of disease assessment. This includes: A \textit{local reader} (gastroenterologist) assigns an initial score, followed by an independent expert's (\textit{central reader}) assessment score. If they disagree, a third \textit{adjudicator expert} provides an independent score. The \textit{final trial} score, used for clinical evaluations, is the median of these three scores.

\subsection{Evidential Deep Learning (EDL) Model}
Deep learning models for UC disease severity assessment often make overconfident predictions, posing risks in ambiguous cases\cite{rubin2023development}. To address this, we implement EDL \cite{sensoy2018evidential}, a method that quantifies uncertainty directly from the model’s output using the Dempster-Shafer Theory (DST). Unlike conventional methods such as MC Dropout or deep ensembles, EDL estimates uncertainty efficiently without requiring multiple models or multiple forward passes.

Our approach processes endoscopy videos from UC trials with $N$ frames by extracting $N \times D$ frame-level features using a foundation model (FM) pre-trained on a large-scale endoscopy dataset. These features are passed to a downstream classifier which comprises of a transformer, Attention-based MIL (ABMIL)\cite{ilse2018attention}, and a dense layer to estimate MES.
Instead of a conventional softmax layer, which only provides point estimates, we introduce EDL by using a Softplus activation and Dirichlet distributions to model both uncertainty and class probabilities. The Softplus function ensures non-negative evidence values, defined as: $e_{ic} = \text{Softplus}(.) = \log(1 + \exp(.))$, where $e_{ic}$ represents the evidence for class $c$ among $C$ classes for a given sample $i$. These evidence values parameterize the Dirichlet distribution, which models a distribution over class probabilities rather than a single-point estimate. The Dirichlet parameters are given by: $ \alpha_i = \langle \alpha_{i1}, \ldots, \alpha_{iC} \rangle$, where $\alpha_i=e_{ic}+1$. Next, the total strength of evidence for sample $i$ is defined as: $S_i = \sum_{c=1}^{C} (e_{ic} + 1)$.

\textbf{Uncertainty and Class Probability Estimation}: In EDL method, the \textbf{uncertainty estimate} $u_i$ is defined as: $u_i = \frac{C}{S_i}$. If no evidence is available ($e_{ic}=0$), the Dirichlet distribution becomes uniform, leading to maximum uncertainty ($u_i \to 1$). This means the model is highly uncertain about its predictions. The \textbf{class probability} for sample $i$ and class $c$ is derived from the mean of the Dirichlet distribution: $p_{ic} = \frac{\alpha_{ic}}{S_i} = \frac{e_{ic} + 1}{S_i}$. 

\textbf{EDL Loss Function}: To ensure well-calibrated predictions, we employ an uncertainty-aware loss function that integrates classification loss with a KL divergence regularization term: 
\begin{gather}
    L = \sum_{i=1}^{N} \sum_{c=1}^{C} y_{ic} \left[ \psi(S_i) - \psi(\alpha_{ic}) \right] + \lambda_t \cdot \sum_{i=1}^{N} KL \left[ D(p_{i} | \alpha_{i}) || D(p_{i} | 1) \right]
\end{gather}
where, $y_{ic}$ is the true label for sample $i$ and class $c$, $\psi(\cdot)$  is the digamma function.
The first term represents classification loss, ensuring correct predictions, and the second term is the KL divergence regularization, preventing overconfident predictions. To balance these losses during training, we apply an annealing coefficient: $\lambda_t = \min(1.0, \frac{t}{T})$, where $t$ is the current epoch and $T$ is a predefined threshold. This allows gradual incorporation of KL loss as training progresses. By integrating EDL, our framework efficiently models both uncertainty and class probabilities in a single forward pass, providing a robust disease assessment.

\subsection{MEGAN: Multi-Expert Gating Network}
MES scoring is inherently subjective, incurring variability in expert annotations. Deep learning models trained on different expert labels and classifier architectures potentially introduce biases and uncertainties. MEGAN addresses this issue by adaptively weighting and aggregating outputs from multiple EDL models.
MEGAN has two variants: MEGAN-Gated employs a gating network (GN) to assign weights for aggregating the probabilities and uncertainties from $K$ EDL models, whereas MEGAN-Naive simply averages the $K$ EDL models' outputs.

\textbf{MEGAN-Gated Architecture}: MEGAN consists of multiple EDL models and a lightweight GN. Each EDL model is trained independently using different expert labels and modeling strategies. With the EDL models kept frozen, the GN takes features from all EDL models and learns to assign optimal weights, combining their predictions into a final score along with an uncertainty estimate.

The GN includes four Multi-Layer Perceptron (MLP) modules:
\textbf{1. Shared MLP}: It processes input features from $K$ EDL models, forming a $K \times d$ matrix ($d$ is the feature dimension) to produce a common shared feature representation for the next three MLPs.
\textbf{2. Probability MLP}: It processes the shared features and applies Tanh activation to generate weights $w^k_{p_i}$ for aggregating model probabilities $p^k_i$ into the final estimate $\hat{p}_i$.
\textbf{3. Uncertainty MLP}: It processes the shared features and applies Sigmoid activation to the shared features to create weights $w^k_{u_i}$ for aggregating uncertainties $u^k_i$ into the final uncertainty estimate.
\textbf{4. Epsilon MLP}: This module refines the final uncertainty $\hat{u}_i$ by adding $\epsilon_i$, which is derived from processing the shared features and using Tanh activation. It allows for flexibility in adjusting the uncertainty score - raising it for incorrect predictions and lowering it for correct ones.

\textbf{GN Training Strategy}:
MEGAN is trained in two stages. First, individual EDL models are trained separately using diverse expert labels and classifier architectures. Next, GN is trained on final MES labels with EDL models frozen, allowing it to learn an optimal weighting strategy without altering EDL outputs.

\textbf{GN Loss Function:}
GN is optimized using a composite loss function:
\begin{gather}
L_{\text{total}} = L_{\text{cls}} + L_{\text{unc}} + L_{\text{eps}} \\
L_{\text{cls}} = -\sum_{i=1}^{N} y_i \log(\hat{p}_i) \\
L_{\text{unc}} = \beta_1 \sum_{i=1}^{N} c_i \cdot \hat{u}_i + \beta_2 \sum_{i=1}^{N} (1 - c_i) \cdot (1 - \hat{u}_i) \\
L_{\text{eps}} = \gamma_1 \sum_{i=1}^{N} c_i \cdot \text{ReLU}(\epsilon_i) + \gamma_2 \sum_{i=1}^{N} (1 - c_i) \cdot \text{ReLU}(-\epsilon_i)
\end{gather}

\textbf{1. Classification Loss} (Eqn. 3) ensures accurate final estimation $\hat{p}_i$. MEGAN minimizes the cross-entropy loss against the final MES label ($y_i$).

\textbf{2. Uncertainty Regularization Loss} (Eqn. 4) encourages GN to assign higher uncertainty to incorrect predictions ($c_i=0$) and lower uncertainty to correct ones ($c_i=1$). $\beta_1, \beta_2$ control penalty strength. This enhances calibration by increasing uncertainty for incorrect predictions.

\textbf{3. Epsilon Regularization Loss} (Eqn. 5) fine-tunes uncertainty estimation.
For correct prediction ($c_i = 1$), negative \( \epsilon_i \) is encouraged to reduce uncertainty. For incorrect prediction ($c_i=0$), positive \( \epsilon_i \) is promoted to increase uncertainty by further refining calibration. Here, $\gamma_1$ and $\gamma_2$ are scaling factors. 

\textbf{Final Probability and Uncertainty}:
After training, GN computes final predictions using weighted aggregation: $\hat{p}_i = \frac{1}{K} \sum_{k=1}^{K} w^k_{p_i} p^k_{i}$. 
Final uncertainty estimate is computed as:
$\hat{u_i} = \frac{\sum_{k=1}^{K} w^k_u u^k_i}{\sum_{k=1}^{K} w^k_u} + \epsilon_i$. This weighting ensures that more reliable models have a greater influence. With these modules, MEGAN effectively integrates multiple EDL models, enhancing MES scoring accuracy, uncertainty estimation, and robustness in clinical trials.

\section{Data Splits, Benchmarking \& Implementation Details}

\hspace{2em}\textbf{Data Splits and Preprocessing}: We follow the dataset partitioning and preprocessing approach from \cite{chaitanya2024arges}. Our study includes endoscopy videos from four clinical trials (two UC \cite{sands2018peficitinib, sands2019ustekinumab}, two CD \cite{sands2022ustekinumab, allez2023phase}) across 30 countries, covering 2,411 patients and 4,911 videos ($\sim$71M frames). The dataset was split into 80\% training and 20\% test sets, with FM pre-trained on the training data. Since MES labels are available only for UC trials, EDL models were trained via 4-fold cross-validation on the UNIFI and JAKUC datasets. Model evaluation was conducted on the held-out test set (20\%), and prospective validation on the unseen QUASAR trial dataset (14M frames) assessed generalization.

\textbf{Model Architectures and Training}: We trained a FM using ViT-B~\cite{dosovitskiy2020image} with DINOv2~\cite{oquab2023dinov2} to extract features for downstream MES scoring, following \cite{chaitanya2024arges}. The baseline Arges model~\cite{chaitanya2024arges} consists of a transformer, an ABMIL, and a dense layer with dropout to estimate MES. It was trained for 20 epochs using a learning rate of $10^{-4}$ and a weight decay of $10^{-5}$.
For EDL models, we integrated a Softplus layer into Arges with a Dirichlet distribution to model class probabilities and optimized it using digamma cross-entropy loss with KL divergence (Eqn. 1). We trained six independent EDL models: four based on \textit{central reader} scores with different architectures with varying number of transformer layers, attention heads, and dropout rates, and two based on \textit{local reader} scores with different architectures.
To address class imbalance, we used weighted sampling. After training the EDL models, we froze their weights and trained a GN with MLP blocks, optimizing it using the composite loss (Eqn. 2) over 20 epochs with a learning rate of $10^{-4}$ and dropout of 0.25. Uncertainty penalties were heuristically set to $\beta_1 = \gamma_1 = 1.0, \beta_2 = \gamma_2 = 5.0$ based on the validation set.

\textbf{Benchmarking and Comparison}:
We benchmarked MEGAN against various uncertainty estimation methods. First, we evaluated the baseline Arges model \cite{chaitanya2024arges} and incorporated MC Dropout, using 40 forward passes at inference to estimate uncertainty, with results aggregated for final scores. Next, we assessed Deep Ensemble models that combined predictions from four independently trained networks.
For uncertainty-aware modeling, we evaluated a single EDL model with a Softplus layer for MES and uncertainty estimation. We also tested Megan-Naive, which averaged predictions from six independent EDL models trained on different expert labels and architectures. Finally, our proposed MEGAN-Gated framework leverages a GN to optimally weight expert models, aggregating predictions from six frozen EDL models.

\textbf{Evaluation Metrics}: 
Performance was evaluated using weighted F1-scores against Final Trial MES labels. Uncertainty calibration was assessed via Expected Calibration Error (ECE): $\text{ECE} = \sum_{m=1}^{M} \frac{|B_m|}{N} \left| \text{acc}(B_m) - \text{conf}(B_m) \right|$, where \( B_m \) is the \( m \)-th confidence bin, \( |B_m| \) is the sample count, $N$ is the total samples, and \( \text{acc}(B_m) \) and \( \text{conf}(B_m) \) denote empirical accuracy and predicted confidence. Lower ECE indicates better calibration.

\section{Experiments, Results and Discussion}

\textbf{4.1. MEGAN Fares Favorably Compared to State-of-the-Art Methods}: Table \ref{table1:combined_f1_and_ece} compares MEGAN (Naive $\&$ Gated) with baselines like MC Dropout, Deep Ensembles, and EDL models across two test sets (UNIFI, JAKUC) and an unseen clinical trial (QUASAR). While Arges-based models achieve reasonable F1 scores, they struggle with calibration (ECE). MC Dropout improves calibration, and ensembles further enhance ECE, consistent with prior work. EDL models, estimating MES scores and uncertainty, outperform baselines in ECE while maintaining strong F1 scores. MEGAN-Naive, leveraging multiple EDL models, improves calibration by 19.4\% and F1 by 1\% over the best baseline. MEGAN-Gated, optimally aggregating EDL models via a gating network, achieves the highest F1 ($+$3.5\%) and lowest ECE ($+$30.5\%) with strong calibration shown in Fig. 2(c) and performs better than MEGAN-Naive. These improvements generalize to QUASAR, underscoring MEGAN’s suitability for clinical deployment. MEGAN-Gated achieves the lowest ECE and higher accuracy (F1), which is crucial in clinical deployments. Also, Gated model provides better class-wise uncertainty for distinguishing confident vs. uncertain cases over the Naive model (Sec. 4.2).

\begin{table}[!t]
  \centering
  {
  \begin{tabular}{|p{2.65cm}|
                  p{0.5cm}|
                  p{0.5cm}|
                  p{0.65cm}|
                  p{1.0cm}|
                  p{1.3cm}|
                  p{1.0cm}|
                  p{1.3cm}|
                  p{1.0cm}|
                  p{1.3cm}|}
    \hline
    \multirow{2}{*}{\textbf{UQ Methods}} & \multirow{2}{*}{ME} & \multirow{2}{*}{UQ} & \multirow{2}{*}{Cost} & \multicolumn{2}{c|}{Test Set UNIFI} & \multicolumn{2}{c|}{Test Set JAKUC} & \multicolumn{2}{c|}{\textbf{Unseen} Set Quasar} \\ 
    \cline{5-10}
    &   &  &  & \textbf{F1} ($\uparrow$) & \textbf{ECE} ($\downarrow$) & \textbf{F1} ($\uparrow$)  & \textbf{ECE} ($\downarrow$) & \textbf{F1} ($\uparrow$)  & \textbf{ECE} ($\downarrow$) \\ \hline
    Arges\cite{chaitanya2024arges} & $\times$ & $\times$ & 1x & 0.614 & 0.302 & 0.521 & 0.354 & 0.654 & 0.221 \\ \hline
    MC Dropout\cite{gal2016dropout} & $\times$ & $\times$ & 40x &0.614 & 0.286 & 0.522 & 0.338 & 0.657 & 0.207 \\ \hline
    Deep ensemble\cite{lakshminarayanan2017simple} & $\times$ & $\times$ & 4x & \underline{0.641} & 0.198 & 0.554 & 0.199 & 0.657 & 0.168 \\ \hline
    EDL\cite{sensoy2018evidential} & $\times$ & $\checkmark$ & 1x & 0.622 & 0.132 & 0.576 & 0.184 & 0.647 & 0.154 \\ \hline
    MEGAN-Naive & $\checkmark$ & $\checkmark$ & 6x & 0.640 & \underline{0.116} & \underline{0.620} & \underline{0.163} & \underline{0.663} & \underline{0.124} \\ \hline
    MEGAN-Gated & $\checkmark$ & $\checkmark$ & 6x & \textbf{0.644 }& \textbf{0.083} & \textbf{0.634} & \textbf{0.144} & \textbf{0.680} & \textbf{0.107} \\ \hline
  \end{tabular}
  }
  \caption{MEGAN-Gated and Naive models consistently outperform baselines in F1-score and calibration score across test sets UNIFI, JAKUC, and the unseen QUASAR set. ME: multiple expert labels used for training models, UQ: uncertainty quantification, and Cost is the compute cost, with x indicating inference on single A100 GPU.}
  \label{table1:combined_f1_and_ece}
\end{table}

\textbf{4.2. Stratifying Samples Using Uncertainty Scores}: Uncertainty scores from EDL, MEGAN-Naive, and MEGAN-Gated enable sample stratification, directing uncertain cases for expert review to reduce workload in large-scale trials. Using class-specific thresholds ($t_c$) dervied from the UNIFI and JAKUC validation sets, QUASAR predictions are classified as confident or uncertain. 
Table 2(a) indicates that confident samples achieve higher accuracy, as shown in the left column, while uncertain samples, flagged for review, exhibit lower accuracy, as illustrated in the right column. MEGAN-Gated outperforms both EDL and MEGAN-Naive in terms of accuracy and retention of confident samples, demonstrating a 10\% increase in retention.
Box plots in Fig. 2(b) highlight MEGAN-Gated’s superior ability to distinguish correct from incorrect predictions via uncertainty scores, except for class 2, which remains ambiguous even for experts. 
Notably, most misclassified samples (red box) exhibit higher uncertainty and correct ones (green box) show lower uncertainty, but this pattern reverses for class 2. 
These results support automated triaging, ensuring accuracy while reducing expert involvement.

\begin{figure}[!t]
    \centering
    \includegraphics[trim=0 0 0 0, clip, width=\textwidth]{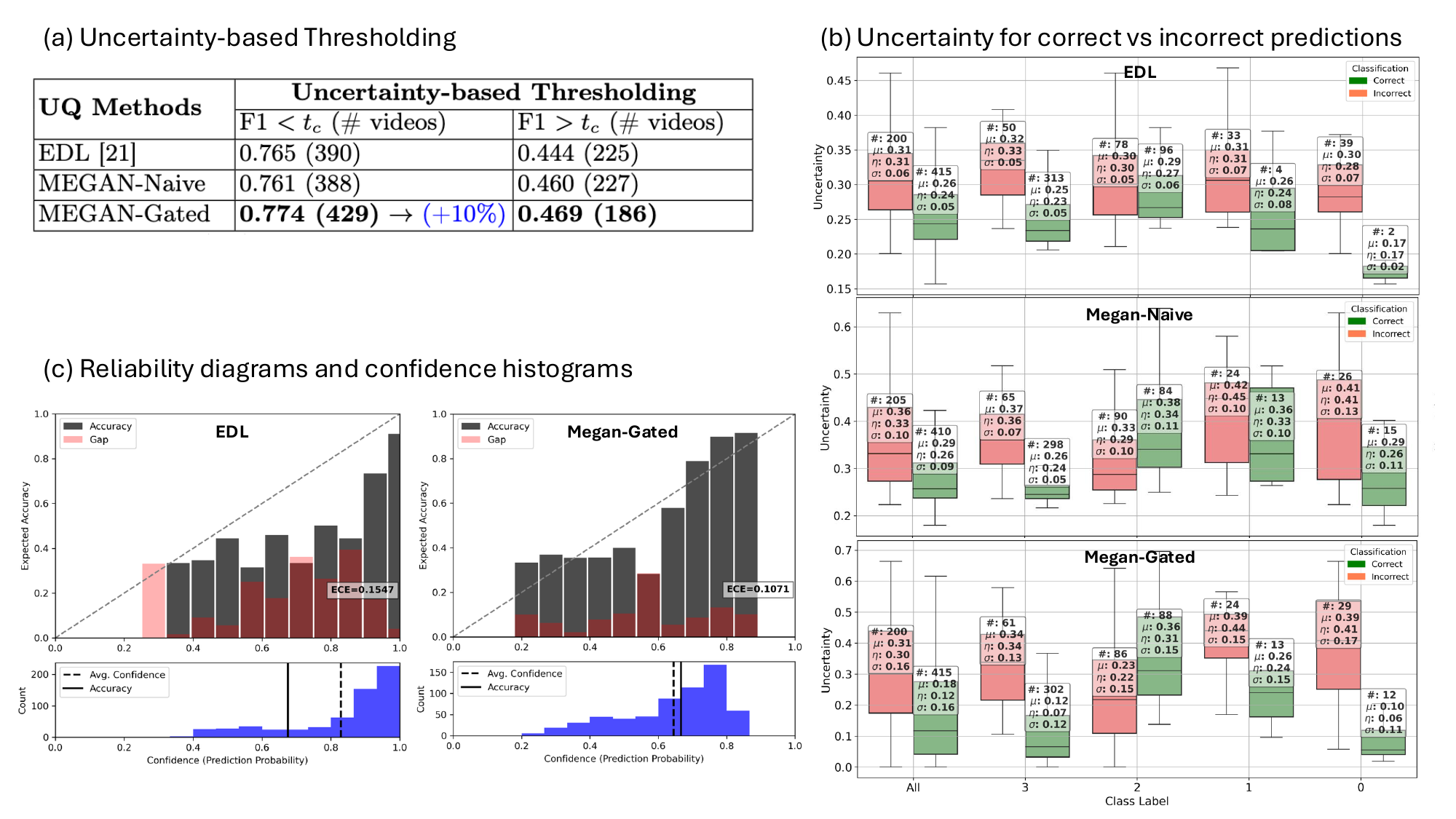}
    \caption{Results on unseen QUASAR data comparing MEGAN-Gated to EDL and MEGAN-Naive: (a) Table displays uncertainty-based results categorized by threshold ($t_c$): the first column shows confident videos below the threshold, while the second column shows uncertain videos above it. This demonstrates that MEGAN can filter a higher number of confident samples with an improved F1 score. (b) Three boxplots illustrate how MEGAN-Gated utilizes uncertainty to distinguish between correct and incorrect predictions across each ground truth class more accurately. (c) Two reliability diagrams and confidence histograms for the unseen QUASAR dataset demonstrate that MEGAN outperforms EDL.}
    \label{fig:uncertainty_uq_quasar}
\end{figure}


\textbf{4.3. Expert Validation of Confident and Uncertain Cases}: Three gastroenterologists reviewed 30 videos from the Quasar set, assigning MES scores (0–3) and confidence ratings (1–5) per video. For confident cases, the consensus expert F1-score compared to final trial labels was 0.61, with an average confidence of 4.34/5, while MEGAN achieved a higher F1-score of 0.66. For uncertain cases, the consensus F1-score dropped to 0.38 with an average confidence of 3.8/5. This drop indicates the difficulty of these cases, which MEGAN successfully identified. In summary, MEGAN provides better predictions for confident cases and effectively identifies difficult cases for further experts' evaluation.


\section{Conclusion}
We introduce MEGAN, a novel multi-expert uncertainty quantification framework for automated UC disease severity assessment. MEGAN incorporates multiple expert labels by training several EDL-based models and optimally aggregating their predictions through a gating network. This approach effectively captures inter-rater variability in endoscopy video assessments while ensuring robust uncertainty estimation. Extensive evaluation on large-scale UC clinical trials demonstrates that MEGAN outperforms existing UQ methods, achieving higher prediction accuracy (F1), better calibration (ECE), and improved uncertainty estimation. By quantifying uncertainty, MEGAN enables targeted expert review, reducing clinical workload. While challenges remain, this work represents a significant step toward capturing multi-expert uncertainty, with potential applications beyond UC in broader clinical decision support systems.
\hfill \break

\textbf{Disclosure of Interests:} All authors were employees of Janssen R\&D, LLC, when conducting this research and may own company stock / stock options.

%
%
%
%
\bibliography{ref}
\bibliographystyle{splncs04}
\end{document}